\documentclass[journal]{IEEEtran}
\usepackage{amsthm,amssymb}
\usepackage{amsmath}
\usepackage{easyReview}
\usepackage{graphicx}
\usepackage{amsmath}
\usepackage{amsfonts}
\usepackage{algorithm}
\usepackage{algorithmic}
\usepackage{makecell}
\usepackage{booktabs}
\setcellgapes{3pt}
\usepackage{multirow}
\setcellgapes{3pt}

\usepackage[justification=centering]{caption}
\usepackage{color}
\usepackage{verbatim}
\usepackage{amssymb}
\usepackage{tabularx}

\usepackage{cite}

\ifCLASSINFOpdf
\else
\fi
\begin{document}
\bstctlcite{IEEEexample:BSTcontrol}
%
\title{CommGPT: A Graph and Retrieval-Augmented Multimodal Communication Foundation Model}

\author{Feibo Jiang, \textit{Senior Member, IEEE}, Wanyun Zhu, Li Dong,  Kezhi Wang, \textit{Senior Member, IEEE}, Kun Yang, \textit{Fellow, IEEE}, Cunhua Pan, \textit{Senior Member, IEEE}, Octavia A. Dobre, \textit{Fellow, IEEE}
	
}

\maketitle
\begin{abstract}
Large Language Models (LLMs) possess human-level cognitive and decision-making capabilities, making them a key technology for 6G.
However, applying LLMs to the communication domain faces three major challenges: 1) Inadequate communication data; 2) Restricted input modalities; and 3) Difficulty in knowledge retrieval. To overcome these issues, we propose CommGPT, a multimodal foundation model designed specifically for communications.
First, we create high-quality pretraining and fine-tuning datasets tailored in communication, enabling the LLM to engage in further pretraining and fine-tuning with communication concepts and knowledge. Then, we design a multimodal encoder to understand and process information from various input modalities.
Next, we construct a Graph and Retrieval-Augmented Generation (GRG) framework, efficiently coupling Knowledge Graph (KG) with Retrieval-Augmented Generation (RAG) for multi-scale learning. 
Finally, we demonstrate the feasibility and effectiveness of the CommGPT through experimental validation.
\end{abstract}


\IEEEpeerreviewmaketitle

\section{Introduction}
The sixth-generation mobile communication technology (6G) aims to build an intelligent world of interconnected devices, offering society an unprecedented information transmission experience. In the 6G architecture, intelligence is one of its key driving forces \cite{9237460}. 6G networks will leverage adaptive resource management, intelligent network optimization, and efficient spectrum utilization to achieve self-learning and self-optimization, allowing dynamic adaptation to complex and changing communication environments in real time.

Large Language Models (LLMs), such as GPT \cite{achiam2023gpt} and LLaMA \cite{touvron2023llama}, represent the cutting edge of generative AI research. With their advanced neural network architectures and training on vast amounts of corpus data, these models exhibit human-level understanding and cognitive abilities. Introducing LLMs into 6G networks will significantly enhance their level of intelligence. This integration will enable the networks to more rapidly perceive, simulate, and respond to the fast-changing communication environment, generalize more efficiently, and understand and adapt to unseen communication scenarios \cite{10700707}. As such, LLMs are poised to become an essential tool for addressing the complex challenges of future wireless communication systems.
\setcounter{subsubsection}{0}
\subsection{Related work}
Several LLMs have already been applied in communication. For instance, TSpec-LLM leverages open-source datasets combined with  Retrieval-Augmented Generation (RAG) to enhance LLMs’ understanding of 3GPP standards, improving their precision in document analysis and response \cite{nikbakht2024tspec}. Tele-LLM is a specialized LLM tailored for telecommunications, leveraging customized datasets to improve its expertise in this field \cite{RN17}. Similarly, TelecomGPT introduces a telecommunications-specific LLM training framework that fine-tunes general LLMs to achieve exceptional performance on telecom-specific tasks \cite{zou2024telecomgpt}. These advancements underscore the significant potential of LLMs in enhancing the intelligence of 6G networks.
\subsection{Challenges of applying LLMs in communication}
Despite their promising performance across many tasks in communication, developing  foundational models suited for the communication field faces several challenges as follows:
\subsubsection{Inadequate communication data}
Mainstream LLMs lack specialized knowledge of communications, limiting their ability to understand and accurately respond to domain-specific concepts and problems. Most LLMs are trained on large-scale general-purpose datasets, while high-quality, domain-specific datasets in communication are relatively scarce. This deficiency restricts LLMs’ ability to provide accurate and reliable responses tailored to the communication field \cite{jiang2024large}.

\subsubsection{Restricted input modalities}
Many standards, protocols, and documents in communication include tables, images, and other non-textual content. Thus, multimodal information is a critical component that communication foundational models need to address. However, current LLMs in communication are primarily designed to process textual data, limiting their ability to understand multimodal content. In scenarios requiring the integration of multiple modalities for comprehensive responses, this limitation can lead to suboptimal performance.

\subsubsection{Difficulty in knowledge retrieval}
 The communication knowledge is highly specialized and complex, requiring retrieval and analysis at different scales. Two primary approaches are currently used for enabling LLMs to access external communication knowledge:
(1) Knowledge Graphs (KGs) enable the organization of complex knowledge networks and facilitate the understanding of communication tasks from a global (large-scale) perspective. (2) RAG tends to focus on a localized (small-scale) understanding and analysis of communication knowledge due to the retrieval of a series of document fragments. However, no approach has been able to effectively balance the demands for both local and global knowledge retrieval\cite{fatehkia2024t}.



\setcounter{subsubsection}{0}
\subsection{Contributions}
To address the aforementioned challenges, we propose CommGPT, a multimodal foundational model in communication, enhanced with multimodal encoder, RAG and KG. Our contributions are summarized as follows:
\subsubsection{High-quality communication dataset}
We construct and release CommData, a high-quality dataset tailored in communication. CommData consists of a pretraining dataset and a fine-tuning dataset. The pretraining dataset includes state-of-the-art communication-related resources, such as recent communication papers, patents, codebases, and 3GPP and IEEE protocol standards. These materials encompass a wide range of topics, including networking technologies, signal processing, and wireless communication protocols, enabling comprehensive knowledge acquisition for LLMs. The fine-tuning dataset is curated and structured to generate question-answer pairs for instruction tuning, improving the LLM's ability to follow communication-specific instructions and accurately understand diverse communication-related commands.

\subsubsection{Multimodal data encoding}
Communication documents often contain tables and images, which traditional LLMs struggle to process effectively. We develop a multimodal encoder that utilizes multimodal embedding models, such as Bootstrapped Language-Image Pretraining (BLIP) \cite{li2023blip} and QOCR \cite{mithe2013optical}, to parse data from various modalities, including images and tables. This design allows the LLM to better understand user intent comprehensively and accurately. By leveraging multimodal encoder, we enhance the accuracy and reliability of communication-related question-answering in multimodal scenarios, reducing hallucinations when handling cross-modal tasks.

\subsubsection{Graph and Retrieval-Augmented Generation (GRG)}
We introduce a multi-scale analysis framework that integrates KG and RAG. This framework creates an efficient pipeline that couples graph knowledge with vector knowledge to improve the accuracy of external communication knowledge retrieval and reasoning. The pipeline employs a fine-tuned LLM to generate a graph-based knowledge base, which is then combined with the vector database in RAG. This enables the LLM to adaptively fuse knowledge for tackling complex problems at different scales, significantly enhancing the completeness and accuracy of communication-related knowledge queries.

The remainder of this paper is organized as follows. In Section II, the key technologies for integrating communication knowledge with foundational models are presented. In Section III, the CommGPT framework and its key components are described. In Section IV, the detailed design of the CommGPT framework is provided. In Section V, experimental results are presented, followed by a discussion of open issues in Section VI, and conclusions are drawn in Section VII.

\setcounter{subsection}{0}
\setcounter{subsubsection}{0}
\section{Learning process of communication foundation model}
For learning communication knowledge, LLMs adopt two primary learning approaches. The first involves pretraining and fine-tuning, embedding communication knowledge directly into the foundation model's parameters. However, this method is time-intensive and unsuitable for accommodating knowledge that requires frequent updates. The second approach integrates RAG and KG, leveraging external vector and graph databases to incorporate communication knowledge into the foundation model’s context learning. This method does not require updates to the foundation model’s parameters, making it better suited for learning rapidly evolving knowledge.
\subsection{Internal learning: pretraining and fine-tuning}
\subsubsection{Unsupervised pretraining}
The objective of pretraining is to equip the foundation model with expert-level communication knowledge, enhancing its ability to comprehend specialized concepts and structured information in communication. First, we load a corpus encompassing knowledge in communication, such as communication protocols, channel models, and communication standards. Next, open-source LLMs, such as LLaMA \cite{touvron2023llama} or Gemma \cite{team2024gemma} are trained on the communication corpus using unsupervised learning tasks, such as masked language modeling and causal language modeling \cite{10766891}. Through these tasks, the LLM learns to predict contextual relationships or fill in missing information. Pretraining enables the foundation model to grasp the underlying structure and patterns of communication knowledge, significantly improving its adaptability to domain-specific terminology and document structures, thereby establishing a solid foundation for subsequent fine-tuning.

\subsubsection{Supervised fine-tuning}
Fine-tuning is an additional optimization step for the pretrained foundation model, focusing on learning from communication data to improve performance on specialized tasks. Through fine-tuning, the LLM can better meet the demands of communication-specific tasks while retaining the general features learned during pretraining. The process begins by restructuring communications datasets to generate instruction datasets tailored to various tasks. The pretrained foundation model is then fine-tuned on these instruction datasets using supervised learning. Fine-tuning typically involves training only a subset of the model's parameters. To avoid catastrophic forgetting, it is essential to design high-quality, high-coverage instruction datasets, which can significantly enhance the LLM's precision and performance in communication-specific tasks.
\subsection{External learning: RAG and KG}
\subsubsection{Vector data-based RAG}
RAG is a method that integrates information retrieval with foundation models to provide accurate contextual knowledge, enhancing both knowledge coverage and response accuracy in complex communication tasks. By retrieving relevant communication knowledge from vector databases before generating outputs, RAG dynamically complements the foundation model’s existing knowledge. This technique addresses challenges related to knowledge updating and detail precision, significantly improving the model’s comprehension and expertise in communication. Moreover, it eliminates the need for frequent large-scale retraining of foundation models to incorporate updated knowledge, offering notable flexibility and scalability. 

\subsubsection{Graph data-based KG}
KGs represent entities and their relationships through a graph structure, systematically capturing and storing communications knowledge via entities  and the relationships between them. This structured knowledge enhances the foundation model's reasoning capabilities and ensures consistency in knowledge representation. Through pretraining and fine-tuning, foundation models can effectively identify entities and explicit relationships in communication, improving the accuracy of KG construction. Furthermore, leveraging KGs enables the LLM to infer implicit relationships between entities, facilitating high-quality relation and event extraction. 
This greatly enhances the LLM's understanding and reasoning capabilities for complex communication knowledge. 
The comparison of pretraining, fine-tuning, RAG, and KG for communication foundation model is illustrated in TABLE \ref{tab:Comp}.

\begin{table*}[ht] 
	\renewcommand{\arraystretch}{1.5} 
	\caption{Comparison of pretraining, fine-tuning, RAG, and KG}
	\label{tab:Comp}
	\begin{center}
		\begin{tabularx}{\textwidth}{|l|X|X|X|X|}
			\hline
			\textbf{Aspect} & \textbf{Pretraining} & \textbf{Fine-tuning} & \textbf{RAG} & \textbf{Knowledge Graph} \\
			\hline
			Goal & Learning general language patterns & Optimizing performance for specific tasks & Retrieving external documents to enhance LLM & Constructing structured knowledge to enhance LLM \\
			\hline
			Data Type & Large-scale unlabeled text & Labeled instruction data & External documents/vector database & Structured graph database \\
			\hline
			Learning Method & Unsupervised learning & Supervised learning & Retrieval and in-context learning & Graph reasoning \\
			\hline
		Tuning	Parameters  & All parameters of LLMs & Task-specific parameters of LLMs & Parameters of retrieval modules & Graph embedding and structure \\
			\hline
			Time Complexity & High & Medium & Low & Medium \\
			\hline
		Disadvantages & Lack of task-specific optimization &  Labeled instruction data requirement & Retrieval suboptimality & Difficulty in generating high-quality graph structure. \\
			\hline
		\end{tabularx}
	\end{center}
\end{table*}

\setcounter{subsection}{0}
\setcounter{subsubsection}{0}
\section{CommGPT framework}
\begin{figure*}[htpb]
	\centering
	\includegraphics[width=17cm]{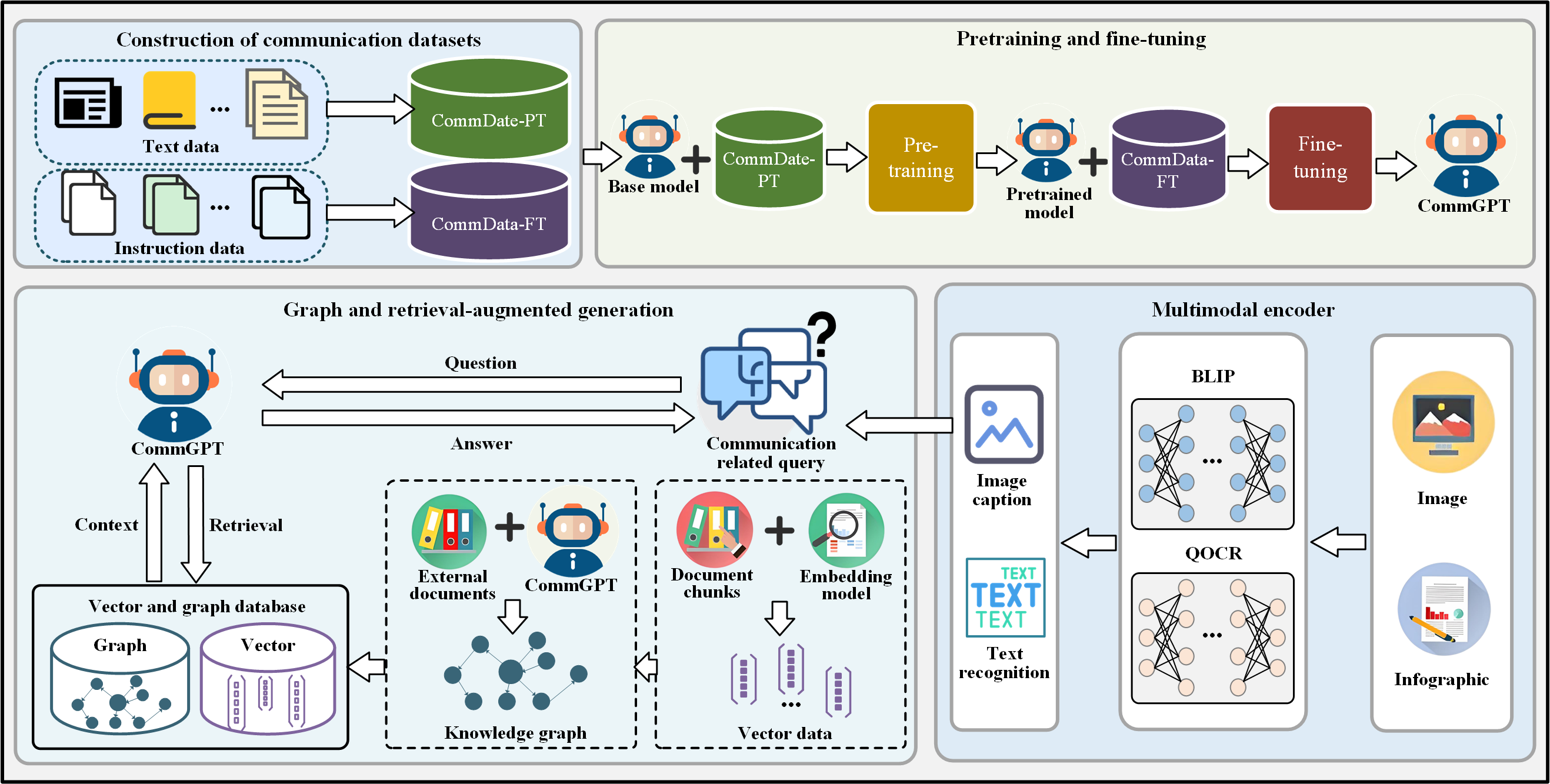}
	\caption{Design framework of the CommGPT system}
	\label{fig:fig1}
\end{figure*}

In this paper, we proposes CommGPT, a graph and retrieval-augmented multimodal communication foundation model. The CommGPT framework establishes an efficient pipeline that combines pretraining, fine-tuning, RAG, and KG to improve the accuracy of the foundation model's understanding, analysis, and decision-making capabilities in communication. The overall design process of the CommGPT framework is illustrated in Fig. \ref{fig:fig1}.

\subsection{Construction of communication datasets}
We develop a communication-specific dataset, CommData, which includes two components: the continuing PreTraining dataset (CommData-PT) and the instruction Fine-Tuning dataset (CommData-FT). The CommData-PT dataset consolidates communication protocols and standards (such as 3GPP and IEEE standards), along with publicly available communication-related data (e.g., Wiki, papers, patents, and codes), enabling the foundation model to learn and internalize communication knowledge. The CommData-FT dataset focuses on optimizing the foundation model for various communication tasks. It includes task-specific instruction data designed around communication protocols and standards, aiming to further enhance the foundation model’s understanding and execution of communication tasks, significantly improving its response accuracy.

\subsubsection{CommData-PT}
To design the CommData-PT dataset, we collected the following types of data:

\begin{itemize}
	\item \emph{3GPP standard data}:
	This dataset contains 15,016 processed 3GPP standard documents (from Rel8 to Rel19).
	
	\item \emph{IEEE standard data}:
	This dataset includes 40 protocol documents related to communications extracted from IEEE standards (such as 802.3, 802.11, 802.15, and C951).
	
	\item \emph{Communication patents}:
This dataset consists of 697,717 communication-related patents spanning from 1975 to 2024.
	
	\item \emph{Communication papers}:
This dataset consists of 90,310 communication-related papers sourced from the Arxiv platform.
	
	\item \emph{Communication code}:
We extract 14,128 code entries related to communications from GitHub. This dataset aims to improve the LLM’s capacity to comprehend and generate communication-related code.
	
	\item \emph{Wiki data}:
	We select 19,543 entries related to communications from Wikipedia, drawn from a total of 6,407,849 records.

\end{itemize}

\begin{figure*}[htpb]
	\centering
	\includegraphics[width=19cm]{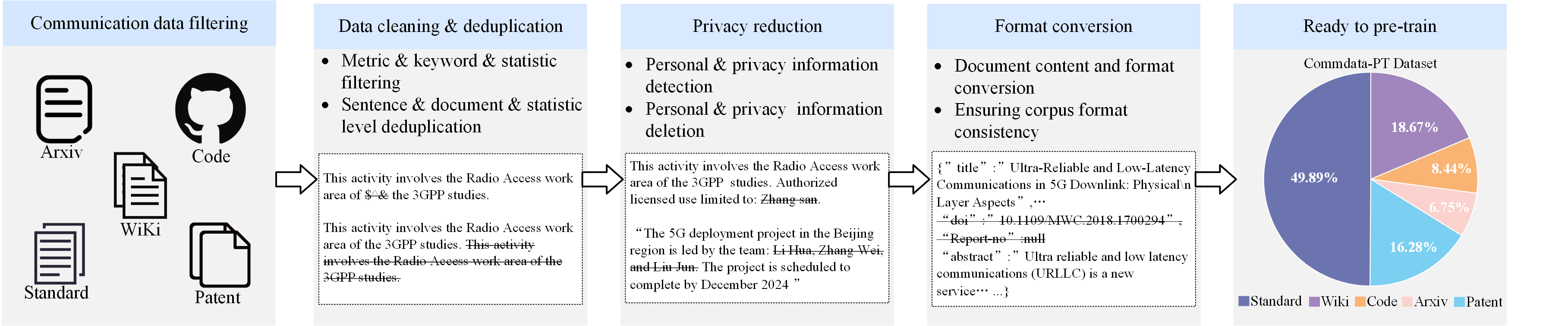}
	\caption{Data preprocessing for CommData-PT}
	\label{fig:fig2}
\end{figure*}

	We employ two methods to extract high-quality communication data from public datasets: (1) \emph{LLM-based Filtering}: Using existing LLMs (e.g., GPT, LLaMA) to identify communication-related documents and filter out irrelevant or low-quality content. (2) \emph{Keyword-based Filtering}: Customizing a set of communication-specific keywords and extracting documents based on these keywords, while removing HTML tags, hyperlinks, and template content as interfering elements. After data collection, preprocessing becomes a crucial step for constructing a high-quality pretraining dataset. The preprocessing steps included eliminating noise, redundancy, irrelevant data, and potentially harmful content to ensure that CommGPT performs optimally. We propose a standardized data preprocessing pipeline for the communications domain, as shown in Fig. \ref{fig:fig2}. 

\subsubsection{CommData-FT}
The goal of CommData-FT dataset is to enable the foundation model to generate expected outputs when given clear task instructions, thereby enhancing its task adaptation, contextual understanding, and accuracy in generating results. A complete communication instruction consists of the following components: \emph{Instruction, Input, Output, and Metadata}. The \emph{Instruction} clearly describes the task; the \emph{Input} provides the contextual information for the instruction; the \emph{Output} offers a high-quality result based on the instruction and input, meeting the task requirements; and the \emph{Metadata} (optional) includes additional details such as task notes or difficulty level.
We use LLaMA 3-8B-Instruct model to generate the CommData-FT dataset based on the 3GPP and IEEE standards, aiming to enhance the foundation model’s understanding and representation of communication knowledge. The specific steps are as follows:

\begin{itemize}
	\item \emph{Generating questions and answers}:
	Using the LLaMA 3-8B-Instruct model, we refine different questions from the CommData-PT dataset based on instructions and guide the model to generate accurate answers.
	
	\item \emph{Building instruction data}:
	The instructions, generated questions, and corresponding answers are paired to form the instruction dataset. Each question is ensured to meet the requirements of the instruction, with accurate and relevant answers.
	
	\item \emph{Quality assessment and filtering}:
	The generated instructions are assessed for quality, and high-quality instructions are selected. This ensures the dataset's diversity and broad coverage of content and task types.
	
\end{itemize}

A specific example of the instruction data is as follows:
\emph{\{"Instruction":” This is a Question and Answer task related to 3GPP.", "Input": "What is the purpose of the SIP-based protocol framework?", "Output": "The SIP-based protocol framework serves as a means of user configuration of supplementary services in the IM CN subsystem.", "Metadata": "Section 4.1, General description in 24238-c00"\} }

In this example, the Instruction is a Question and Answer task, the Input and Output are generated from the LLaMA 3-8B-Instruct model, and the Metadata is the data source.


\vspace{-2pt}
\subsection{Pretraining and fine-tuning}
Using CommData, we design a two-stage training scheme. First, we perform continued pretraining on the base model to enhance its understanding of foundational concepts and unique structures in the communication domain. Then, we apply instruction fine-tuning to further improve its performance on specific communication tasks. This combination of communication knowledge adaptation and task optimization ensures that the base model strikes a balance between broad communication knowledge and specialized expertise in various tasks.

\subsubsection{Continued pretraining}
We perform  continued pretraining to further train a foundation model on the CommData-PT dataset. This method allows the foundation model to retain general knowledge while acquiring specialized communication knowledge. We use open-source LLMs (such as LLaMA, Gemma, etc.) as the foundation model, and employ an autoregressive architecture for unsupervised training. The self-attention mechanism helps the foundation model capture long-range dependencies and complex contextual information. By predicting the probability of each token in its context, the foundation model acquires expertise in communication knowledge, improving its understanding of communication concepts and standards. After the pretraining phase, the foundation model is evaluated to ensure the learning quality in communication. 

\subsubsection{Instruction fine-tuning}
We further fine-tune the pretrained foundation model using the CommData-FT dataset to optimize its performance on communication tasks. This step enables the foundation model to adapt to specific tasks such as answering questions, multiple-choice questions, and generating codes. We perform supervised efficient fine-tuning on the pretrained foundation model, employing the Low-Rank Adaptation (LoRA) \cite{zheng2024llamafactory}. The LoRA parameters are adjusted through backpropagation, and task-specific loss functions such as cross-entropy loss are used for optimization. Fine-tuning typically uses a smaller learning rate to avoid disrupting the communication knowledge acquired during pretraining. Finally, the fine-tuned foundation model is saved in preparation for subsequent tasks.

\subsection{Multimodal encoder}
To enable the understanding of tables and images in communication documents, we integrate two multimodal encoders into the input layer of the foundation model, where the BLIP encoder is used to extract high-level semantic features from images \cite{li2023blip}, while the QOCR encoder extracts lower-level text information in infographics \cite{mithe2013optical}, thereby providing comprehensive support for multimodal data in communication documents.

\subsubsection{BLIP encoder}
BLIP is a generative model designed for image semantic understanding. It captures high-level features from images using a self-attention mechanism and converts them into high-quality textual descriptions, enabling textual encoding of images in communication documents. The process of generating image descriptions with the BLIP encoder is as follows: First, BLIP uses a vision encoder based on the Vision Transformer architecture to extract high-level semantic features from the image. Then, through the image-text fusion module, it aligns the image and text features, allowing BLIP to better understand the textual meaning associated with the image. Finally, BLIP generates descriptive text for the image using a Transformer-based autoregressive generator in the text decoder \cite{li2023blip}.

\subsubsection{QOCR encoder}
QOCR is a deep learning model for detecting and recognizing text information in infographics. It converts the image files into grayscale by performing scaling and color space transformations. Next, Convolutional Neural Networks (CNNs) are employed to extract features from the image and identify text regions within it. Then, Long Short-term Memory Networks (LSTMs) are used to model the sequences in the text regions, precisely mapping the character sequences extracted from the image into readable text. Finally, a confidence threshold is applied to filter the recognition results, discarding those with confidence levels below the set threshold, ensuring that only high-confidence recognition results are retained, thereby enhancing the overall accuracy and reliability of the output \cite{mithe2013optical}.

\subsection{Graph and retrieval-augmented generation}
To efficiently learn and apply knowledge in communication, we propose a multi-scale learning framework that integrates KG and RAG. We use the fine-tuned foundation model to generate a KG of communication corpora, creating a global structured representation of entities and their relationships in communication documents. Simultaneously, we employ RAG to chunk communication documents, build a vector database, and enable local retrieval of knowledge from these documents. The combination of these two approaches enables the foundation model to efficiently acquire and generate knowledge in various communication task scenarios, demonstrating exceptional adaptability and accuracy. The workflow of the GRG process is as follows:

\subsubsection{Building the vector database}
After preprocessing the input communication documents, they are first chunked into smaller units. Each document chunk corresponds to an independent unit of information, such as a paragraph or technical concept. The document chunks are then encoded into high-dimensional vector embeddings using an embedding model and stored in a vector database. These vectorized representations retain the deep semantic features of the document chunks, making them suitable for large-scale document retrieval. A vector index is constructed in the database to facilitate fast retrieval of the most relevant document chunks by comparing query vectors with stored vectors.

\subsubsection{Building the graph database}
Using CommGPT, entities, attributes, and relationships in the communication documents are identified. By aligning entities, we construct a relationship network between different entities and attributes, forming a structured KG. In this KG, nodes represent entities, and edges represent the relationships between entities. The Neo4j graph database is used to store and manage the KG, supporting efficient graph queries \cite{9085182}. This allows the foundation model to analyze and process the complex network of communication entities and relationships, further enhancing its global understanding of specialized knowledge in communication.

\subsubsection{Joint retrieval of vectors and graphs}
The image encoder interprets the semantic content of the image and converts it into the text information needed for the query when the user's input contains image information. Then, the foundation model converts the user's query into a query vector and retrieves the most relevant document chunks from the vector database. The model compares the query vector with stored vectors to select the document chunk whose corresponding vector is the most similar to the query vector. The result of the vector retrieval represents the local knowledge relevant to the query.
Subsequently, the foundation model extracts key entities from the user's query and performs reasoning based on the relationships between entities in the graph database. It retrieves the attributes of these entities and their relationships with other entities. The result of the graph retrieval represents the global knowledge related to the query.

\subsubsection{Context-augmented generation}
The relevant document chunks retrieved from the vector database and the entity attributes and relationships inferred from the graph database are merged to form enriched contextual information, which is input into the fine-tuned CommGPT. At this stage, CommGPT is able to understand the terminology and technical background of the communications, reducing hallucinations and generating more accurate responses. For instance, when answering questions that involve multiple entities and technical relationships, the CommGPT is able to clearly display the connections between entities and provide accurate technical interpretations.

\begin{figure}[htbp]
	\centering
	\includegraphics[width=9cm]{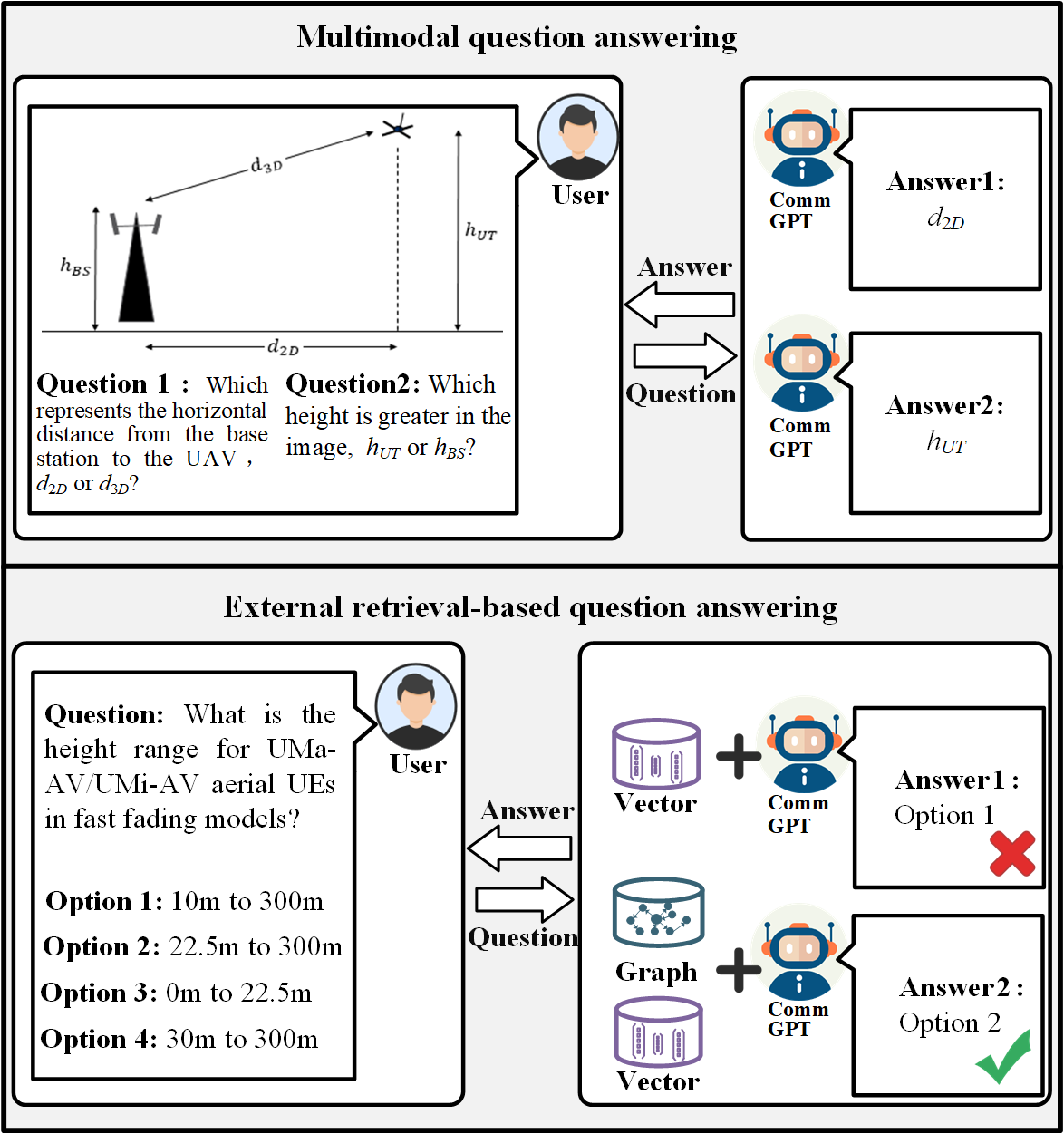}
	\caption{Examples of CommGPT applied to 3GPP queries.}
	\label{fig:fig3}
\end{figure}

Examples of CommGPT applied to 3GPP queries are presented in Fig. \ref{fig:fig3}. 

\vspace{-10pt} 

\setcounter{subsection}{0}
\setcounter{subsubsection}{0}
\section{Experiments}
\subsection{Experimental setup}
During the training phase, CommGPT utilizes the open-source Gemma 2-9b-instruct model \cite{team2024gemma} from Google as its base model, which is then pretrained and fine-tuned on the CommData dataset. The learning rate scheduler is set to cosine annealing, with the initial learning rate of the pretraining phase set to $5 \times 10^{-6}$ and the fine-tuning phase set to $1 \times 10^{-5}$. The optimizer is set to Adam. The rank for LoRA is set to 8, with a scaling factor of 16. Mixed-precision training with BF16 is employed.
Moreover, BLIP is used as the high-level image encoder, QOCR as the low-level image encoder, QAnything as the RAG system, and Milvus as the vector database. The KG is generated using GraphRAG in combination with CommGPT, and Neo4j is used as the graph database. 3GPP\_TR \cite{nikbakht2024tspec} dataset, which includes questions of three difficulty levels (easy, intermediate, and hard), is applied to evaluate the CommGPT and its competitors. The accuracy is defined as the ratio of correctly answered questions by LLMs to the total number of questions in 3GPP\_TR.


\subsection{Ablation experiment}
The competitors used in this experiment are as follows:

\subsubsection{BaseModel (Gemma 2-9b)}
The open-source foundation model with no pretraining or fine-tuning on CommData.

\subsubsection{CommGPT}
The communication foundation model pretrained and fine-tuned using CommData.

\subsubsection{CommGPT-R}
The CommGPT enhanced with RAG.

\subsubsection{CommGPT-GRG}
The CommGPT enhanced with both RAG and KG.


\begin{figure}[htbp]
	\centering
	\includegraphics[width=9cm]{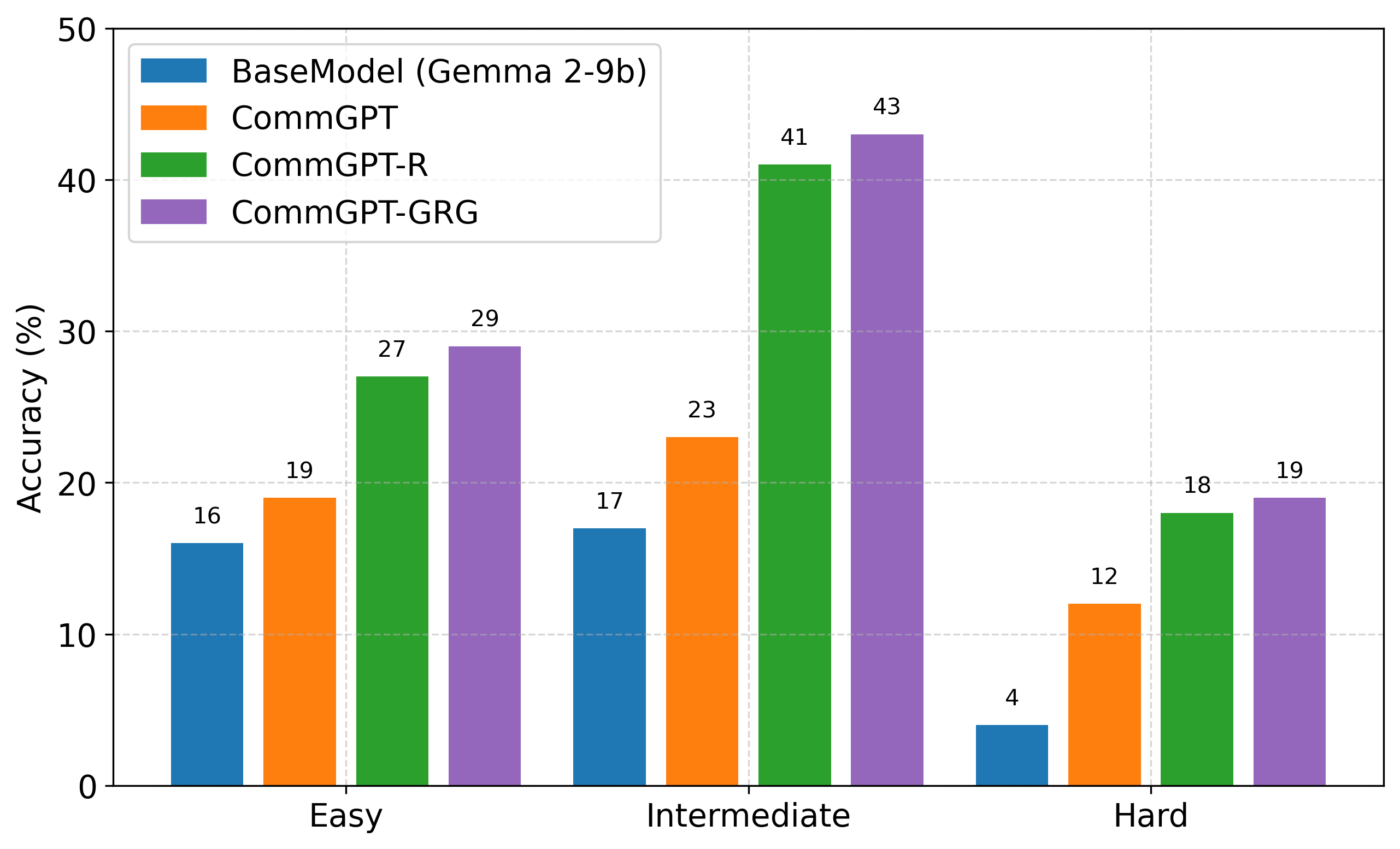}
	\caption{The accuracy of LLMs in the ablation experiment.}
	\label{fig:fig4}
\end{figure}

The accuracy for each LLM is shown in Fig. \ref{fig:fig4}. The experimental results demonstrate a significant improvement in the accuracy of CommGPT for communication question-answering tasks. Initially, the accuracy of the BaseModel is only 37\%. After pretraining and fine-tuning on the CommData dataset, the accuracy of CommGPT increases to 54\%. This indicates that fine-tuning the BaseModel on a specialized communication dataset enables it to learn features and patterns more closely aligned with communication knowledge, thereby improving its ability to handle communication question-answering tasks. Furthermore, after integrating KG with RAG, the accuracy of the CommGPT-GRG model increases further to 91\%. This highlights that the GRG can provide multi-scale external knowledge retrieval, effectively supplementing the CommGPT’s inherent communication knowledge deficiencies and greatly enhancing its contextual understanding capability.

\subsection{Comparative experiment}
The competitors used in this experiment are as follows:
\subsubsection{LLaMA2-7B Series}
LLaMA 2-7B and LLaMA 2-7B-Instruct are second-generation open-source LLMs proposed by Meta, with 7 billion parameters.

\subsubsection{LLaMA3-8B Series}
LLaMA 3-8B and LLaMA 3-8B-Instruct are third-generation open-source LLMs proposed by Meta, with 8 billion parameters.

\subsubsection{GPT Series}
GPT-3.5 and GPT-4 are closed-source LLMs developed by OpenAI.

\subsubsection{Gemini 1.0}
A closed-source multimodal LLM developed by Google.

\subsubsection{Tele-LLM}
An LLM optimized for telecommunications, developed by Yale University, with 8 billion parameters \cite{RN17}.

\subsubsection{Tspec-LLM Series}
GPT-3.5-T, Gemini 1.0-T, and GPT-4-T are telecom-specific LLMs based on GPT-3.5, Gemini 1.0, and GPT-4, respectively, enhanced with the RAG \cite{nikbakht2024tspec}.



\begin{figure}[htbp]
	\centering
	\includegraphics[width=9cm]{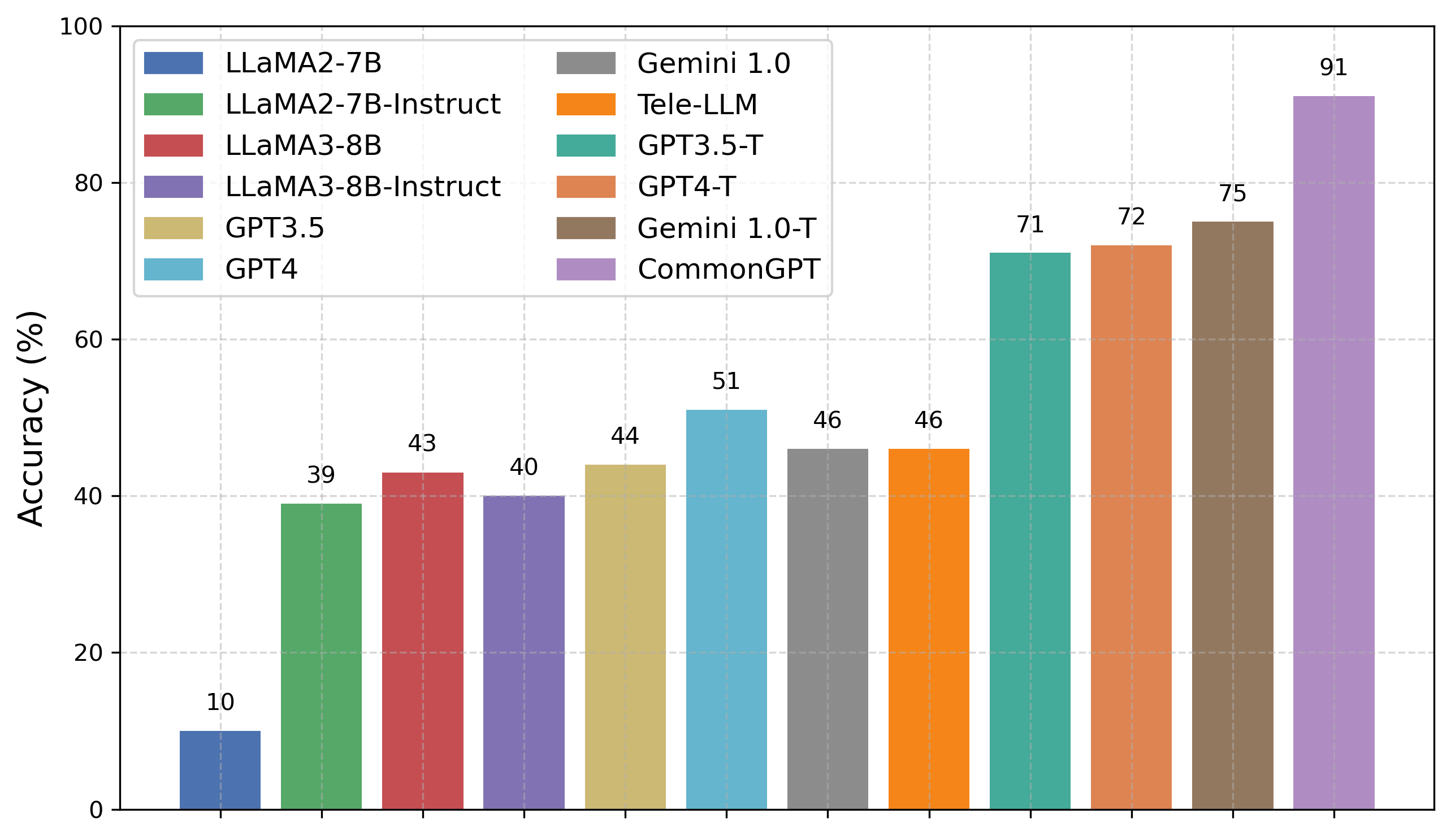}
	\caption{The accuracy of LLMs in the comparative experiment.}
	\label{fig:fig5}
\end{figure}

The detailed experimental results are presented in Fig. \ref{fig:fig5}, which illustrates that the accuracy of open-source LLMs (e.g., the LLaMA2 and LLaMA3 series) is the lowest, while closed-source LLMs (e.g., the GPT series and Gemini 1.0) exhibit higher accuracy, with GPT-4 achieving the highest performance. Among the three groups of telecom-specific LLMs, the open-source Tele-LLM demonstrated performance comparable to that of the closed-source models, suggesting that training on specialized telecom data enhances the performance of open-source LLMs in communication. Furthermore, the accuracy of the Tspec-LLM series (e.g., GPT-3.5-T, Gemini 1.0-T, and GPT-4-T), which integrates RAG, shows significant improvement, indicating that RAG technology, by incorporating external telecom knowledge, enhances the LLM's ability to address telecom-related questions. Finally, CommGPT, which utilizes the GRG, achieved the highest accuracy, highlighting that the combination of KG and RAG facilitates a better understanding of domain-specific terminology and complex issues in communication, thereby providing more accurate and efficient responses.

\section{Conclusion}
In this paper, we proposed a multimodal foundational model for communications, named CommGPT. First, we introduced CommData, a specially curated dataset for CommGPT, which encompasses various types of data, including protocols, standards, papers, patents, and codes. Then, we employed CommData to pretrain and fine-tune the foundational model. Additionally, we designed multimodal encoders to enhance the CommGPT's ability to handle multimodal data in communication documents. Finally, we proposed a multi-scale learning approach that combines KG and RAG, assisting CommGPT in generating more precise and comprehensive responses for communications. Experimental results demonstrated that CommGPT outperforms other competitors in communication question-answering tasks, exhibiting the highest accuracy.

\bibliographystyle{IEEEtran}
\bibliography{EndNote}

\begin{thebibliography}{10}
\providecommand{\url}[1]{#1}
\csname url@samestyle\endcsname
\providecommand{\newblock}{\relax}
\providecommand{\bibinfo}[2]{#2}
\providecommand{\BIBentrySTDinterwordspacing}{\spaceskip=0pt\relax}
\providecommand{\BIBentryALTinterwordstretchfactor}{4}
\providecommand{\BIBentryALTinterwordspacing}{\spaceskip=\fontdimen2\font plus
\BIBentryALTinterwordstretchfactor\fontdimen3\font minus
  \fontdimen4\font\relax}
\providecommand{\BIBforeignlanguage}[2]{{%
\expandafter\ifx\csname l@#1\endcsname\relax
\typeout{** WARNING: IEEEtran.bst: No hyphenation pattern has been}%
\typeout{** loaded for the language `#1'. Using the pattern for}%
\typeout{** the default language instead.}%
\else
\language=\csname l@#1\endcsname
\fi
#2}}
\providecommand{\BIBdecl}{\relax}
\BIBdecl
\renewcommand{\BIBentryALTinterwordstretchfactor}{4}

\bibitem{9237460}
H.~Yang \emph{et~al.}, ``Artificial-intelligence-enabled intelligent 6g
  networks,'' \emph{IEEE Network}, vol.~34, no.~6, pp. 272--280, 2020.

\bibitem{achiam2023gpt}
J.~Achiam \emph{et~al.}, ``{GPT-4} technical report,'' \emph{arXiv preprint
  arXiv:2303.08774}, 2023.

\bibitem{touvron2023llama}
H.~Touvron \emph{et~al.}, ``Llama: Open and efficient foundation language
  models,'' \emph{arXiv preprint arXiv:2302.13971}, 2023.

\bibitem{10700707}
S.~Long \emph{et~al.}, ``6g comprehensive intelligence: network operations and
  optimization based on large language models,'' \emph{IEEE Network}, pp. 1--1,
  2024.

\bibitem{nikbakht2024tspec}
R.~Nikbakht, M.~Benzaghta, and G.~Geraci, ``Tspec-llm: An open-source dataset
  for llm understanding of 3gpp specifications,'' \emph{arXiv preprint
  arXiv:2406.01768}, 2024.

\bibitem{RN17}
A.~Maatouk \emph{et~al.}, ``Tele-llms: A series of specialized large language
  models for telecommunications,'' \emph{arXiv preprint arXiv:2409.05314},
  2024.

\bibitem{zou2024telecomgpt}
H.~Zou \emph{et~al.}, ``Telecomgpt: A framework to build telecom-specfic large
  language models,'' \emph{arXiv preprint arXiv:2407.09424}, 2024.

\bibitem{jiang2024large}
F.~Jiang \emph{et~al.}, ``Large language model enhanced multi-agent systems for
  6g communications,'' \emph{IEEE Wireless Communications}, 2024.

\bibitem{fatehkia2024t}
M.~Fatehkia, J.~K. Lucas, and S.~Chawla, ``T-rag: lessons from the llm
  trenches,'' \emph{arXiv preprint arXiv:2402.07483}, 2024.

\bibitem{li2023blip}
J.~Li \emph{et~al.}, ``Blip-2: Bootstrapping language-image pre-training with
  frozen image encoders and large language models,'' in \emph{International
  conference on machine learning}.\hskip 1em plus 0.5em minus 0.4em\relax PMLR,
  2023, pp. 19\,730--19\,742.

\bibitem{mithe2013optical}
R.~Mithe, S.~Indalkar, and N.~Divekar, ``Optical character recognition,''
  \emph{International journal of recent technology and engineering (IJRTE)},
  vol.~2, no.~1, pp. 72--75, 2013.

\bibitem{team2024gemma}
G.~Team \emph{et~al.}, ``Gemma 2: Improving open language models at a practical
  size,'' \emph{arXiv preprint arXiv:2408.00118}, 2024.

\bibitem{10766891}
V.~Mazzia \emph{et~al.}, ``A survey on knowledge editing of neural networks,''
  \emph{IEEE Transactions on Neural Networks and Learning Systems}, pp. 1--17,
  2024.

\bibitem{zheng2024llamafactory}
Y.~Zheng \emph{et~al.}, ``Llamafactory: Unified efficient fine-tuning of 100+
  language models,'' \emph{arXiv preprint arXiv:2403.13372}, 2024.

\bibitem{9085182}
Y.~Zou and Y.~Liu, ``The implementation knowledge graph of air crash data based
  on neo4j,'' in \emph{2020 IEEE 4th Information Technology, Networking,
  Electronic and Automation Control Conference (ITNEC)}, vol.~1, 2020, pp.
  1699--1702.

\end{thebibliography}

\section*{Biographies}
\textbf{Feibo Jiang} (jiangfb@hunnu.edu.cn) is currently an Associate Professor at Hunan Normal University, China.

\textbf{Wanyun Zhu} (zhuwanyun@hunnu.edu.cn) is currently pursuing the master’s degree at Hunan Normal University, China.

\textbf{Li Dong} (Dlj2017@hunnu.edu.cn) is currently an Associate Professor at Hunan University of Technology and Business, China.

\textbf{Kezhi Wang} (Kezhi.Wang@brunel.ac.uk) is a Professor with the Department of Computer Science, Brunel University London, U.K.

\textbf{Kun Yang} (kyang@ieee.org) 
is currently a Chair Professor in the School of Intelligent Software and Engineering, Nanjing University, China.

\textbf{Cunhua Pan} (cpan@seu.edu.cn) is currently a full professor in Southeast University, China. 

\textbf{Octavia A. Dobre} (odobre@mun.ca) is a professor and Canada Research Chair
Tier 1 at Memorial University, Canada.


\end{document}